\documentstyle[preprint,eqsecnum,aps,epsf]{revtex}
\newcommand{\be}{\begin{equation}}
\newcommand{\ee}{\end{equation}}
\newcommand{\ba}{\begin{eqnarray}}
\newcommand{\ea}{\end{eqnarray}}
\begin{document}
\title{
${}^{~~~~~~~~~~~~~~~~~~~~~~~~~{published ~in ~the}}$\\
${}^{\it{Advances ~in ~Space ~Research, ~Vol. ~29, ~p. ~725 ~- ~734~(2002)}}$\\
${}^{}$\\
${}^{}$\\
EULER, JACOBI, AND MISSIONS TO COMETS AND ASTEROIDS.
${}^{}$\\
{\small{Invited Lecture, 33rd COSPAR               
Scientific Assembly, Warsaw, Poland, 16-23 July 2000}}
${}^{}$\\
}
\author{{\large{\bf{Michael Efroimsky}}\\}}
\address{Institute for Mathematics \& its Applications, ~University of Minnesota}
\address{207 Church Street SE, Suite 400, Minneapolis MN 55455 USA} 
\address{e-mail: efroimsk@ima.umn.edu\\
${}^{}$\\}
\maketitle
\begin{abstract}
Whenever a freely spinning body is found in a complex rotational state, this 
means that either the body experienced some interaction within its 
relaxation-time span, or that it was recently ``prepared'' in a non-principal 
state. Both options are encountered in astronomy where a wobbling rotator is 
either a recent victim of an impact or a tidal interaction, or is a fragment of
a disrupted progenitor. Another factor (relevant for comets) is outgassing. By 
now, the optical and radar observational programmes have disclosed that complex
rotation is hardly a rare phenomenon among the small bodies. Due to impacts, 
tidal forces and outgassing, the asteroidal and cometary precession must be a 
generic phenomenon: while some rotators are in the state of visible tumbling, a
much larger amount of objects must be performing narrow-cone precession not so 
easily observable from the Earth.

The internal dissipation in a freely precessing top leads to relaxation 
(gradual damping of the precession) and sometimes to spontaneous changes in the
rotation axis. Recently 
developed theory of dissipative precession of a rigid body reveals that this is
a highly nonlinear process: while the body is precessing at an angular rate $
\omega$, the precession-caused stresses and strains in the body contain components 
oscillating at other frequencies. Dependent upon the spin state, those frequencies 
may be higher or, most remarkably, lower than the precession rate. In many states 
dissipation at the harmonics is comparable to or even exceeds that at the principal 
frequency.

For this and other reasons, in many spin states the damping of asteroidal and 
cometary wobble happens faster, by several orders, than believed previously. 
This makes it possible to measure the precession-damping rate. The narrowing of
the precession cone through the period of about a year can be registered by the
currently available spacecraft-based observational means. We propose an 
appropriate observational scheme that could be accomplished by comet and
asteroid-aimed missions. Improved understanding of damping of excited rotation 
will directly enhance understanding of the current distribution of small-body 
spin states. It also will constrain the structure and composition of excited 
rotators.

However, in the near-separatrix spin states a precessing rotator can 
considerably slow down its relaxation. This lingering effect is similar to the 
one discovered in 1968 by Russian spacecraft engineers who studied free wobble
of a tank with viscous fuel.
\end{abstract}

\pagebreak

\section{~~Prolegomena~~~~~~~~~~~~~~~~~~~~~~~~~~~~~~~~~~~~~~~~~~~~~~~~~~~~~~~~~~~~~~~~~~~~~~~~~~~~~~~~~~~~~~~~~}

In 1730, 23-year-old leutenant Leonhard Euler retired from the Russian navy, to
become a professor of physics at the Russian Academy of Sciences. Eleven years 
later he interrupted his tenure, to assume the post of Director of Mathematics 
and Physics at the Berlin Academy, offered to him by King Friedrich the Second.
There he stayed until 1765 when he was invited back to St.Petersburg by 
the enlightened Empress Catherine the Great. It is during that 25-year-long 
Berlin period of his life that Euler made his major contributions to the 
mechanics of a rotating body. After publishing some prefatory results in 1750 
and 1758, Euler wrote down, in 1760, his now celebrated equations describing 
free spin of an unsupported top with arbitrary moments of inertia $\;I_1,\;I_2,
\;I_3\;$:
\be
I_i \; {\dot{\Omega}}_i \; - \; \left(I_j \; - \; I_k \protect\right) \; 
\Omega_j \; \Omega_k \; = \; \tau_i \; \; \; .
\label{1.1}
\ee
$\;\Omega_{1,2,3}\;$ being the the angular-velocity components in the 
coordinate system defined by the three principal axis of inertia, (1, 2, 3).
For a freely rotating body, the external torques $\;\tau_{1,2,3}\;$ standing in
the right-hand side are nil. Without loss of generality, one may assume that $
\;I_3\,\geq\,I_2\,\geq\,I_1\;$, and thus, always to consider axis (3) to be the
major-inertia axis.

This result was published only five years later, in Chapter X of Euler's book 
{\textit {Theoria motus corporum solidorum seu rigidorum ex primis nostrae 
cognitionis principiis stabilita et ad omnes motus qui in huiusmodi corpora 
cadere possunt accomodata.}}

In Chapter XI, Euler easily found the solution for a prolate symmetrical 
case\footnote{Later the symmetric top was also treated by 
Lagrange (1788) and Poisson (1813).}: if $\;I_3\,=\,I_2\,>\,I_1\;$ (like, for 
example, in the case of an elongated rod), the vector of inertial angular 
velocity $\;\bf \Omega\;$ describes a circular cone about the minor-inertia 
axis (1) of the body:
\be
{\Omega}_1 \; \; = \; \; const\;\;,\;\;\;
{\Omega}_2 \; \; = \; \; {\Omega}_{\perp} \cos {\omega}t~~,~~~
{\Omega}_3 \; \; = \; \; {\Omega}_{\perp} \sin {\omega}t~~,~~~
\label{1.2}
\ee
$\omega =(I_1/I_3  - 1) \Omega_1 $ being the precession rate. It is 
possible to show that the angular-momentum vector $\,\bf J\,$ behaves  
like $\,\bf \Omega\,$, i.e., precesses about axis (1) at the same angular rate. 
This picture of  $\,\bf \Omega\,$ and  $\,\bf J\,$ 
precessing about axis (1) is the one seen by an observer associated with the 
body frame. In an inertial observer's opinion, things will look
different: from his viewpoint the angular momentum $\,\bf J\,$ of an 
unsupported top will conserve, while its angular velocity $\,\bf \Omega\,$ and 
the minor-inertia axis (1) will be precessing about $\,\bf J$.

Similarly, in the case of oblate symmetry ($\;I_3\,>\,I_2\,=\,I_1\;$) both 
$\;\bf \Omega\;$ and $\;\bf J\;$ will perform, in the body frame, a circular 
precession about the major-inertia axis (3):
\be
{\Omega}_1 \; \; = \; \; {\Omega}_{\perp} \cos {\omega}t~~,~~~
{\Omega}_2 \; \; = \; \; {\Omega}_{\perp} \sin {\omega}t~~,~~~
{\Omega}_3 \; \; = \; \; const
\label{1.3}
\ee
where $\,\omega = (I_3/I_1 - 1) \Omega_3.\;$ In an inertial frame, though, it will be $\;\bf \Omega\;$ and axis (3) describing circular
cones about $\;\bf J\;$

In Chapter XIII Euler tackled the general case of $I_3
>I_2 \geq I_1$ and solved it in terms of functions presently 
known as elliptic integrals. These were pioneered a century 
earlier by John Wallis and Isaac Newton, and went in the late XVIII - 
early XIX centuries under the name of elliptic functions\footnote{See, for 
example, http://www-groups.dcs.st-andrews.ac.uk/~history/HistTopics/}. (Euler, 
though, used neither of these names, and did not refer to Wallis or 
Newton.) Nowadays the name ``elliptic'' belongs to functions 
{\textit {sn, cn, dn}} and their kin, that were not known at the time of Euler.
They were introduced by Karl Jacobi in 1829 (Jacobi 1829), studied by Legendre (1837), and  
later employed (Jacobi 1849, 1882) in the rotating-top studies. These functions are, in a 
way, generalisations of our customary 
trigonometric functions: while for symmetric prolate and oblate 
bodies the circular precession is expressed by (\ref{1.2}) and (\ref{1.3}) 
correspondingly, in the general case $I_3 \geq I_2 \geq I_1$ the 
solution will read: 
\begin{eqnarray}
\Omega_1\;=\;\gamma\;\,{\it{dn}}\left(\omega t , \; k^2 \protect\right)\;\;,
\;\;\;\;\Omega_2 \; = \; \beta \, \; sn\left(\omega t , \; k^2 \protect\right)
\;\;,\;\;\;\;\Omega_3 \; = \; \alpha\;\,{\it{cn}}\left(\omega t,\;k^2
\protect\right)\;\;,\;\;\;
\label{1.4}
\end{eqnarray}
for $\;{\bf{J}}^2 \; < \; 2\;I_2 \; T_{\small{kin}} \; $, and  
\begin{eqnarray}
\Omega_1\;=\;{\tilde \gamma}\;\,{\it{cn}}\left({\tilde \omega} t,\;{\tilde k}^2
\protect\right)\;\;,\;\;\;\;\Omega_2 \; = \;{\tilde \beta}\,\;{\it sn}\left(
{\tilde \omega} t ,\;{\tilde k}^2 \protect\right)\;\;,\;\;\;\;\Omega_3 \; = \; 
{ \alpha}\;\,{\it{dn}}\left({\tilde \omega} t,\;{\tilde k}^2
\protect\right)
\label{1.5}
\end{eqnarray}
for $\;\;{\bf{J}}^2\;>\;2\;I_2\;T_{\small{kin}}\;$. Similarity between 
(\ref{1.2}) and (\ref{1.4}), as well as between (\ref{1.3}) and (\ref{1.5}), is
evident. In the above expressions, the precession rate $\;\it \omega\;$ and the
parameters $\;\alpha ,\;\beta ,\;{\tilde \beta}, \;\gamma ,\;{\tilde \gamma}, 
\;{\tilde \omega},\;k\;$ and $\;{\tilde k}\;$ are some sophisticated 
combinations of $\;I_{1,2,3}, \;T_{\small {kin}}\;$ and $\;{\bf J}^2\;$.
What is important, is that solution (\ref{1.4}) approaches (\ref{1.2}) in the 
limit of prolate symmetry, $\;(I_3\,-\,I_2)/I_1\,\rightarrow\,0\;$, while 
solution (\ref{1.5}) approaches (\ref{1.3}) in the limit of oblate symmetry, 
$\;(I_2\,-\,I_1)/I_1\,\rightarrow\,0\;$. To adumbrate in an illustrative 
manner the applicability realms of (\ref{1.4}) and (\ref{1.5}), let us turn to 
Figure 1. In the course of free spin, two quantities (integrals of 
motion) are conserved. One is the angular momentum 
\be
{\bf{J}}^2 \;=\;I_1^2\,{\Omega}_1^2\;+\;I_2^2\,{\Omega}_2^2\;+\;I_3^2\,
{\Omega}_3^2 \;\;,\;\;\;
\label{1.6}
\ee
another is the kinetic energy
\be
T_{\small{kin}}\;=\;\frac{1}{2}\;\left\{I_1\,{\Omega}_1^2\;+\;I_2\,{\Omega}_2^2
\;+\;I_3\,{\Omega}_3^2 \right\}\;\;.\;\;\;
\label{1.7}
\ee
Evidently, these two expressions define ellipsoids in the angular-velocity 
space $\;(\Omega_1,\,\Omega_2,\,\Omega_3)\;$. Intersection of these two 
surfaces will be the trajectory described by vector $\;\bf \Omega\;$ in the 
said space. On Figure 1, the angular-momentum ellipsoid is depicted. On its 
surface, we have marked the lines of its intersection with several different 
kinetic-energy ellipsoids appropriate to different energies. It does not take 
much space imagination to understand that, for a fixed angular-momentum surface
(\ref{1.6}), there exist an infinite family of kinetic-energy surfaces 
(\ref{1.7}) intersecting with it. The largest surface of kinetic energy 
(corresponding to the maximal value of $\;T_{kin}\;$) is an ellipsoid that
fully encloses our angular-momentum ellipsoid and only touches it in point A 
and its opposite. Similarly, the smallest surface of kinetic energy 
(corresponding to minimal $\;
T_{kin}\;$) would be an ellipsoid fully contained inside our angular-momentum 
ellipsoid and only touching it from inside, at point C and its opposite. It is easy to 
demonstrate that, for a fixed $\;\bf J\;$, the maximal and minimal possible 
values of the kinetic energy are achieved during rotations about the 
minimal-inertia and maximal-inertia axes, correspondingly. It can also be 
shown, from ($\ref{1.1}$), that
in the case of a non-dissipative and torque-free rotation, the tip of

\pagebreak
\break
\begin{figure}
\centerline{\epsfxsize=3.5in\epsfbox{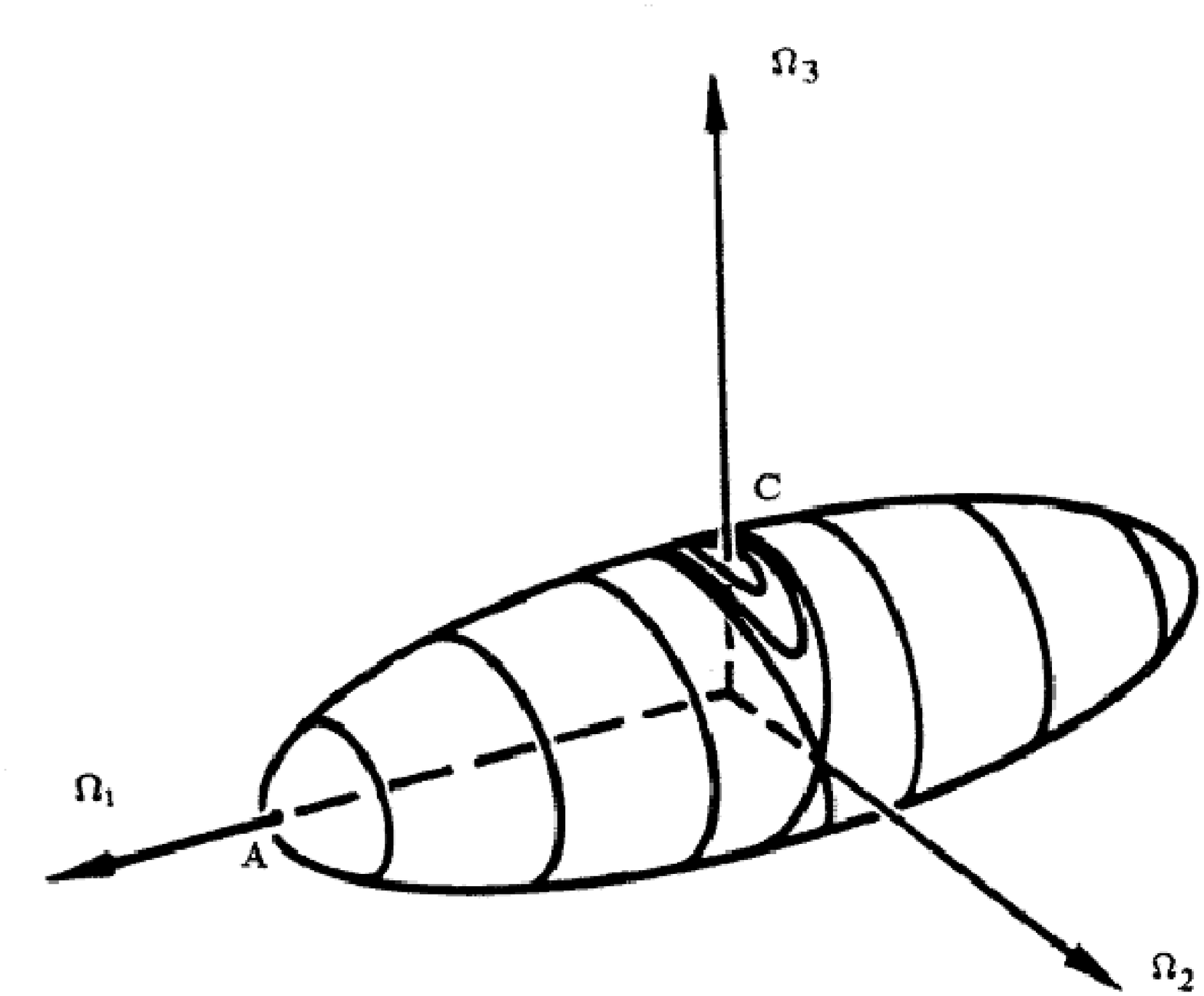}}
\bigskip
\caption{The constant-angular-momentum ellipsoid, 
in the angular-velocity space. The 
lines on its surface are its intersections with the kinetic-energy ellipsoids 
corresponding to different values of the rotational energy. The quasi-stable 
pole A is the maximal-energy configuration, i.e., the state wherein the body 
spins
about its minimal-inertia axis. The stable pole C symbolises the minimal-energy
state, i.e., rotation about the maximal-inertia axis. The angular-velocity 
vector describes the constant-energy lines, and at the same time slowly 
shifts from one line to another, approaching pole C. The picture illustrates 
the case of an elongated body: $I_3 \stackrel{>}{\sim}I_2>I_1$. The 
trajectories are circular near A and remain (in the case of an elongated body) 
virtually circular almost up to the separatrix. The trajectories 
will regain a circular shape only in the closemost proximity of C.}
\end{figure}
\pagebreak
\noindent
the 
vector $\;{\bf{\Omega}}\;$ will be describing, on Figure 1, a curve along
which the angular-momentum and energy ellipsoids intersect (Lamy \& Burns 
1972). Hence, these intersections may be called trajectories. 
Solution ($\ref{1.4}$) is valid for higher energies, i.e., from pole A 
through the separatrix; solution ($\ref{1.5}$) works for lower energies, i.e., 
from the separatrix through pole C. Wherever the trajectories are almost 
circular, the solutions ($\ref{1.4}$) and ($\ref{1.5}$) may be approximated by 
($\ref{1.2}$) and ($\ref{1.3}$), correspondingly.

The formalism developed by Euler and refined by Jacobi might be a perfect tool
for description of rotation of asteroids, comets, cosmic-dust granules, 
spacecrafts and whatever other unsupported rigid rotators, if not for 
one circumstance, inner dissipation. Because of this circumstance, the 
Euler-Jacobi theory of precession works only for time spans
short enough to neglect dissipation. The presence of inner dissipation may be 
guessed even on a heuristic level. The bounded range of permissible energies 
makes one think that a freely spinning body of a fixed angular momentum must 
be seeking ways of relaxation, i.e., of getting rid of the excessive energy, in
order to approach the minimal-energy configuration. Thence the necessity of 
some dissipation mechanism. 

Two such mechanisms are known. One is relevant only for mesoscopic 
rotators, like interstellar-dust grains, and therefore plays a certain role in 
the cosmic-dust alignment. This is the Barnett dissipation, a phenomenon called
into being by the periodic remagnetisation of a precessing paramagnetic body 
(Lazarian \& Draine 1997).

The second mechanism, inelastic dissipation, is, too, relevant for mesoscopic 
grains (Lazarian \& Efroimsky 1999), and it plays the decisive role in the
macroscopic bodies' relaxation. The effect results from the alternating 
stresses produced inside a wobbling body by the time-dependent acceleration of 
its parts. The stresses deform the body, and the inelastic effects cause 
dissipation of the rotational energy.

The dissipation entails relaxation of the precession: the major-inertia axis of
the body and its angular-velocity vector $\;\bf \Omega \;$ tend to align along
the angular momentum $\bf J$. In other words, the precession cone described by
$\;\bf \Omega \;$ about $\bf J$ will be narrowing until $\;\bf \Omega \;$ 
aligns along $\bf J$ completely. A simple calculation (Efroimsky 2001, 
Efroimsky 2000, Efroimsky \& Lazarian 2000, Lazarian \& Efroimsky 1999)
shows that in this case the major-inertia axis of the body will align in the 
same direction, so that, from the body-frame viewpoint, $\;\bf \Omega \;$ will 
eventually 
be pointing along this axis. This configuration will correspond to the minimal 
kinetic energy, the angular momentum being fixed. 

An inertial observer will thus see the unsupported body miraculously changing 
its rotation axis. This is exactly what happened in 1958 when, to mission 
experts' surprise, satellite Explorer I changed its rotation axis. The 
spacecraft was a very elongated body. It had been supposed to spin about 
its least-inertia axis (i.e., about its longest dimension), but 
refused to do so, and instead started precessing (Thomson 1961).

This was probably the first example of a practical need for a further 
development of the Eulerian theory of a free top, a development that would 
address an unsupported top with dissipation. Another motivation for this work 
was put forward in the same year by Prendergast (1958) who studied the asteroid
population of the Solar System and enquired as to how many asteroids could be 
in non-principal (i.e., precessing) spin states, and how this could 
evidence of the impact frequency in the main belt. (Prendergast 
implied that it is collisions that might drive asteroids out of the principal
state and make them wobble.) An important point made by Prendergast was the
generation of the second harmonic: if a body is precessing at an angular rate
$\;\omega\;$, then the dissipation is taking place not only at this frequency
but also at double thereof. Prendergast failed to notice the emergence of the 
higher harmonics, but even his noticing of the second harmonic was an important
observation. In several other aspects the mathematical treatment of the 
problem, offered by Prendergast, was erroneous and gave him no chance to come 
to a reasonable solution. Moreover, at that time the observational astronomy 
lacked any reliable data on wobbling asteroids. So, Prendergast's paper was 
forgotten (even though once in a while it appeared in the references), and his 
successors had to start up from scratch.

The interest in the asteroidal precession re-emerged in 70-s, after the 
publication of the important work (Burns \& Safronov 1973) that suggested 
estimates for the relaxation time, based on the decomposition of the 
deformation pattern into bulge flexing and bending, and also on the conjecture 
that ``the centrifugal bulge and its associated strains wobble back and forth 
relative to the body as the rotation axis {\bf $\; \bf \omega\;$} moves through
the body during a wobble period.'' As turned out later, the latter conjecture 
does not work, because the inelastic dissipation, for the most part of it, is 
taking place not near the surface but in the depth of the body, i.e., not right
under the bulge but deep beneath it. Thus, the bulge is much like an iceberg 
tip. This became clear when the distribution of precession-caused stresses was 
calculated, with improved boundary conditions (Efroimsky \& Lazarian 2000), 
(Lazarian \& Efroimsky 1999). Another, main, problem of Burns \& Safronov's 
treatment was their neglection of the nonlinearity, i.e., of the second and 
higher harmonics. The nonlinearity, in fact, is essential. Neglection thereof 
leads to a large underestimation of the damping rate, because the leading 
effect comes often from the second and higher harmonics
(Efroimsky \& Lazarian 2000), (Efroimsky 2000). All in all, the 
neglection of nonlinearity and mishandling of the boundary conditions leads to 
a several-order underestimate of the precession-damping rate.

In the same year, Peale published an article dealing with inelastic relaxation
of nearly spherical bodies (Peale 1973), and there he did take the second 
harmonic into account.

In 1979 Purcell addressed a similar problem of interstellar-grain precession  
damping. He ignored the harmonics and mishandled the
boundary conditions upon stresses: in (Purcell 1979) the normal stresses had 
their maximal values on the free surfaces and vanished in the centre of the 
body (instead of being maximal in the centre and vanishing on the surfaces). 
These oversights lead to a several-order underevaluation of the dissipation 
effectiveness and, thereby, of the relaxation rate.

\section{~~Precession ~damping~~~~~~~~~~~~~~~~~~~~~~~~~~~~~~~~~~~~~~~~~~~~~~~~~~~~~~~~~~~~~~~~~~~}

The dynamics of precession relaxation is described by the angular rate of
alignment of the maximal-inertia axis (3) along the angular momentum $\bf J$, 
i.e., by the decrease in angle $\;\theta\;$ between these. In the case of 
oblate symmetry (when $\;I_3\;>\;I_2\;=\;I_1\;$), this angle remains 
adiabatically 
unchanged over the precession period, which makes $d{\theta}/dt$ a perfect 
measure of the damping rate (Efroimsky \& Lazarian 2000). However, in the 
general case of a triaxial body angle $\;\theta\;$ evolves periodically through 
the precession cycle. To be more exact, it evolves $\it{almost}$ periodically, 
and its value at the end of the cycle is only slightly different from that in 
the beginning of the cycle. The relaxation is taking place through accumulation
of these slight variations over many periods. This is called adiabatic 
regime, i.e., regime with two different time scales: we have a ``fast'' process
(precession) and a ``slow'' process (relaxation). Under the adiabaticity 
assumption, one may average  $\;\theta\;$, or some function thereof, over the 
precession cycle. Then the damping rate will be described by the evolution of 
this average. Technically, it is convenient to use the average of its squared 
sine  (Efroimsky 2000). One can write for a triaxial rotator:
\be
\frac{d\;<\sin^2 \theta >}{dt}\;=\;\frac{d\;<\sin^2\theta>}{dT_{kin}}\;\;
\frac{dT_{kin}}{dt}\;\;\;,\;\;\;\;
\label{2.1}
\ee
while for an oblate one the expression will look simpler:
\be
\left(\frac{d\, \theta }{dt}\right)_{(oblate)}\;=\;\left(\frac{d\,\theta}{
dT_{kin}}\right)_{(oblate)}\;\;\frac{dT_{kin}}{dt}\;\;\;.\;\;\;\;
\label{2.2}
\ee
The derivatives $\;d\;<\sin^2\theta>/dT_{kin}\;$ and 
$\;\left(d\,\theta/dT_{kin}\right)_{(oblate)}\;$ appearing in (\ref{2.1}) and 
(\ref{2.2}) indicate how the rotational-energy dissipation affects the value of
$\;<\sin^2 \theta>\;$ (or simply of $\;\theta\;$, in the oblate case). These 
derivatives can be calculated from the equations of motion (see Efroimsky \& 
Lazarian (2000) and Efroimsky (2000)). The kinetic-energy decrease, $\;
dT_{kin}/dt\;$, is caused by the inelastic dissipation:  
\be
d{T}_{kin}/dt \; = \; < d{W}/dt > \;\; \;,\;\;\;
\label{2.3}
\ee
$W\;$ being the energy of the alternating stresses, and $\;<...>\;$ 
denoting an average over a precession cycle. (This averaging is justified 
within our adiabatic approach.) Finally, in the general case of a triaxial top,
the alignment rate will read:
\be
\frac{d\,<\sin^2\theta>}{dt}\;=\;\frac{d\,<\sin^2\theta>}{dT_{kin}}\;\;
\frac{d\,<W>}{dt}\;\;\;,\;\;\;
\label{2.4}
\ee
and for a symmetrical oblate top:
\be
\left(\frac{d\, \theta }{dt}\right)_{(oblate)}\;=\;\left(\frac{d\,\theta}{
dT_{kin}}\right)_{(oblate)}\;\;\frac{d\,<W>}{dt}\;\;\;.\;\;\;\;
\label{2.5}
\ee
Vibrations in the material, at frequency $\;\omega\;$, are accompanied by 
dissipation 
\begin{eqnarray}
\frac{dW(\omega)}{dt}\;=\;\frac{2\;\omega\;\,<W(\omega)>}{Q(\omega)}
\label{1.8}
\end{eqnarray}
$\;<W(\omega)>\;$ being the averaged-over-cycle elastic energy of deformation
at frequency $\;\omega\;$, and $\;Q(\omega)\;$ being the so-called quality 
factor of the body material. To calculate $\;<W(\omega)>\;$, one has first to 
find the time-averaged elastic-energy density, $\;d\;<W(\omega)>/dV\;$ (that 
can be calculated from the known picture of stresses and strains) and then to 
integrate it over the volume of the body. This yields:
\begin{eqnarray}
\frac{dW(\omega)}{dt}\;=\;\frac{2\;\omega}{Q(\omega)}\;\int\;\left(\frac{
d\;\,<W(\omega)>}{dV}\right)\;dV\;\;.\;\;
\label{1.9}
\end{eqnarray}
In a more refined approach, though, one should take into account the generation
of harmonics and the inhomogeneity of the wobbling top: 
\begin{eqnarray}
\frac{dW}{dt}\;=\;\sum_{n}\;2\;\omega_n\;\int\;\frac{1}{Q(\omega_n)}\;\left(
\frac{d\;\,<W(\omega_n)>}{dV}\right)\;dV\;\;\;\;
\label{1.10}
\end{eqnarray}
where it is implied that $\;Q\;$ depends not only upon the frequency but also 
upon the coordinates inside the body. The Q-factor is empirically introduced in
seismology and acoustics in order to make up for our inability to describe the 
total effect of several attenuation mechanisms (Nowick and Berry 1972), (Burns 
1986), (Burns 1977), (Knopoff 1963). These bear a complicated dependence upon 
frequency, but the overall frequency dependence of $Q$ is usually very slow and
smooth 
(except some narrow resonances revealing a local domination of one or another 
particular mechanism). The temperature dependence of $Q$ is inseparably 
connected with the frequency dependence (see (Efroimsky \& Lazarian 2000) and 
references therein). The dependence of $Q$ upon the humidity (and upon the 
presence of some other saturants) still poses a challenge to material 
scientists. In many minerals, for example in silicate rocks, several monolayers
of water may decrease $Q$ by a factor of about 55 as described, for example, by
Tittman, Ahlberg, and Curnow (1976) who studied samples of the lunar 
rock\footnote{The absence of moisture in the lunar rock (except, perhaps, in 
some local spots) is the reason that the moonquake echo propagates for long 
with almost no attenuation. It would be dissipated much faster, should the 
lunar rocks contain just a tiny amount of water.}. Presumably, humidity and 
other saturants affect the inter-grain interactions in rocks. 

\section{~~The ~Origin ~of ~the ~Nonlinearity~~~~~~~~~~~~~~~~~~~~~~~~~~~~~~~~~~~~~~~~~~~~~~~}

What is important about formulae (\ref{2.4}), (\ref{2.5}), is that, after 
(\ref{1.10}) is plugged in, one can explicitly see the contributions to the
entire effect, coming from the principal frequency $\,\omega_1\,$ and from the 
harmonics $\,\omega_n\,\equiv\,n\,\omega_1\,$. When vector $\bf{\Omega}$ 
describes approximately circular trajectories on Figure 1, the principal frequency
$\,\omega_1\,$ virtually coincides with the precession rate $\,\omega\,$. 
This doesn't hold, though, when $\bf{\Omega}$ get closer to the separatrix: there $\,\omega_1
\,$ becomes {\it{lower}} than the precession rate. The analysis of 
the stress and strain distributions, and the resulting expressions for $\,d\,
<W>/dt\,$ written down in (Efroimsky \& Lazarian 2000) and (Efroimsky 
2000) shows that the nonlinearity is essential, in that the
generation of harmonics is not a high-order effect but a phenomenon playing a 
key role in the relaxation process. In other words, dissipation associated with
the harmonics is often of the same order as that at the principal 
frequency. Near the separatrix it may be even higher.

The nonlinearity emerges due to the simple fact that the acceleration of a 
point within a wobbling object contains centrifugal terms that are quadratic in
the angular velocity $\;{\bf{\Omega}}\;$. In neglect of small terms caused by 
the body deformation, the acceleration will read:
\be
{\bf{a}}\;\;=\;\;{\bf{{\dot{\Omega}}}}\;\times\;{\bf{r}}\;+ \; 
{\bf{\Omega}} \; \times \; ( {\bf{\Omega}} \times {\bf{r}}) \; \; \; \; .
\label{2.9}
\ee
${\bf{a}} \,$ being the acceleration in the inertial frame, and  $\, {\bf{r}}
\,$ being the position of a point. In the simpliest case of oblate symmetry, 
the 
body-frame-related components of the angular velocity are expressed by 
(\ref{1.3}) plugging whereof into (\ref{2.9}) produces terms containing 
$\,\sin \omega t \,$ and  $\,\cos  \omega t \,$, as well as those containing 
$\,\sin 2  \omega t \,$ and  $\,\cos 2  \omega t \,$. The alternating stresses 
and strains caused by this acceleration are linear functions of $\,\bf a\,$ and, thus, will 
also contain the second harmonic, along with the principal frequency. Calculation of the stresses, strains, and
of the appropriate elastic energy $\,W\,$ is then only a matter of some 
elaborate technique. This technique (presented in (Efroimsky \& Lazarian 
2000) and (Efroimsky 2001)) leads to an expression for $\,W\,$, with 
contributions from $\,\omega\,$ and $\,2 \omega\,$ explicitly separated.
The nonlinearity is essential: in many rotation states the $\,2 \omega$ input in                     
(\ref{1.10}) and (\ref{2.4}) is of order and even exceeds that coming from the principal
frequency $\,\omega$. To explain in brief the reason why the nonlinearity is
strong, we would mention that while the acceleration and the stresses and strains 
are quadratic in the (precessing) angular velocity $\,\Omega\,$, the elastic energy
is proportional to the product of stress and strain tensors. Hence the elastic energy 
is proportional to the fourth power of $\,\Omega\,$.
                                                                             
\section{~~The ~Near-Separatrix ~Slowing-Down ~of ~the ~Precession\\
(Lingering ~Effect)~~~~~~~~~~~~~~~~~~~~~~~~~~~~~~~~~~~~~~~~~~~~~~~~~~~~~~~~~~~~~~}                 
                                                                              
In the general case of a triaxial rotator, precession 
is described by (\ref{1.4}) or (\ref{1.5}). The acceleration of a point inside 
the body (and, therefore, the stresses and strains in the material) will,     
according to (\ref{2.9}), contain terms quadratic in the Jacobi functions.    
These functions can be decomposed in converging series (the so-called nome    
expansions) over $\sin$'es and $\cos$'ines of $\,n \nu$, $n\,$ being odd      
integers for $\;{\it{sn( \omega t,\,k^2)}}\;$ and $\;{\it{cn( \omega t,\,k^2)}}
\;$  and even integers for $\;{\it{dn( \omega t,\,k^2)}}\;$. Here $\;\nu\;$ is 
a frequency {\it{lower}} than the precession rate $\;\omega\,$:              
\be                                                                           
\nu\;=\;\omega\;\frac{2\,\pi}{4\,K(k^2)} \;\;\;\;\;,\;\;\;                    
\label{3.2}                                                                   
\ee                                                                           
$4K(k^2)$ being the mutual period of ${\it{sn}}$ and $                        
{\it{cn}}$. Near the poles $\nu \rightarrow \omega $,                         
while on approach to the separatrix $ \nu \rightarrow 0$. When two such      
expansions get multiplied by one another, they produce a series               
containing all sorts of products like $(\sin\,$m$\nu t\;\sin\,$n$\nu t)\;$, $\; 
(\cos\,$m$\nu t\;\cos\,$n$\nu t)$, and cross terms. Hence the                 
acceleration, stress and strain contain the entire multitude of overtones. 
Even though the further averaging of $W$ over the precession cycle weeds
out much of these terms, we are eventually left with all the harmonics on our hands. 

As explained in (Efroimsky 2000), higher-than-second harmonics will bring
only high-order contributions to the precession-relaxation process when the 
rotation state is described by a point close to poles A or C. Put differently, 
it is sufficient to take into account only the frequencies $\nu \approx \omega
$ and $2 \nu \approx 2 \omega $, insofar as the trajectories on 
Figure 1 are approximately circular (i.e., when (\ref{1.4}) and (\ref{1.5}) are 
well approximated by (\ref{1.2}) and (\ref{1.3})). Near the separatrix the 
situation is drastically 
different, in that all the harmonics become important. We 
thus transit from the domain of essential nonlinearity into the regime of 
extreme nonlinearity, regime where the higher harmonics bring more in the 
process than $\nu$ or $2 \nu $. We are reminded, however, that en route
to the separatrix we not just get all the multiples of the principal frequency,
but we face a change of the principal frequency itself: according to 
(\ref{3.2}), the principal frequency $ \nu $ will be lower than the 
precession rate $ \omega $! This regime may be called 
 ``exotic nonlinearity''. 

Without getting bogged down in rigorous mathematics (to be attended to 
in a separate paper), let me just mention here that in the limit of 
$\bf{\Omega}$ approaching the separatrix the dissipation rate will vanish,
in the adiabatic approximation. This may be guessed even from the fact that in 
the said limit $\nu \rightarrow 0$. We thus come to an important conclusion 
that the relaxation rate, being very high at a distance from the separatrix, 
decreases in its closemost vicinity. Can we, though, trust that the relaxation
rate completely vanishes on the separatrix? No, because in the limit of $\bf{
\Omega}$ approaching the separatrix the adiabatic approximation will fail. 
In other words, it will not be legitimate to average the energy 
dissipation over the precession cycle, because near the separatrix the 
precession rate will not necessarily be faster than the relaxation rate. A 
direct calculation shows that even on the separatrix itself the acceleration
of a point within the body will remain finite (but will, of course, vanish at 
the unstable middle-inertia pole). The same can be said about stress,
strain and the relaxation rate. So what we eventually get is not a 
near-separatrix trap but just lingering: one should expect relaxing tops to
considerably linger near the separatrix. As for Explorer, it is now 
understandable why it easily went wobbling but did not rush to the 
minimal-energy spin state: it couldn't cross the separatrix so quickly. We would call 
it "lingering effect". There is nothing mysterious in it. The capability of 
near-intermediate-axis spin states to mimic simple rotation was pointed out by 
Samarasinha, Mueller \& Belton (1999) with regard to comet Hale-Bopp. A similar setting 
was considered by Chernous'ko (1968) who studied free precession of a tank filled with 
viscous liquid and proved that in that case the separatrix is crossed within a finite 
time interval\footnote{Such problems have been long known also to mathematicians 
studying the motion with a non-Hamiltonian perturbation: the perturbation wants the 
system to cross the separatrix, but is not guaranteed to succeed in it, because
some trajectories converge towards the unstable pole (Neishtadt 1980)} .

\section{Confirmation from NEAR's Observations of 433 Eros}

A calculation 
(Efroimsky 2001) shows that the relaxation rate exponentially slows down 
at small residual angles: the narrower the precession cone the slower the
relaxation rate. This means that many small bodies in the Solar System should 
have retained some residual narrow-cone precession. Their precession may be
too narrow to be seen from the Earth, even by a radar. Still, a close
monitoring from a spacecraft should reveal at least a weak residual wobble.
This is especially the case for Eros: first, because it is a monolith rock and, second, because
it is at the stage of leaving 
the main belt and is a Mars-crosser (so that its rotation was, most probably, 
sometimes disrupted by tidal forces or impacts through the past several millions 
of years). 

Another good reason for Eros to retain some wobble is its almost perfect dynamic 
prolatness ($\,I_3\,\approx\,I_2\,\gg\,I_1$), prolateness that makes 
pole C on Figure 1 so tightly embraced by the separatrices. Even a
weak interaction will push the tip of vector $\,\bf \Omega\,$ across the separatrix 
towards pole A (Black et al 1999), while its relaxation-caused return to C will be slowed down by the 
aforementioned lingering effect. 

Despite all these circumstances, Eros was found in a relaxed or almost\footnote{The 
word "almost" means that the half-angle of the precession cone does not exceed 0.1 
angular degree. (Andrew Cheng, private communication.)} relaxed spin state
(Yeomans 2000). This means that the inelastic relaxation is indeed a very effective process,
as predicted by our study. It is far more effective than believed previously\footnote{To reconcile 
the observations with the old theory (Burns \& Safronov 1973), one would have to hypothesise that 
433 Eros has experienced no impact or tidal disruptions for many (dozens or hundreds) millions of years,
which is very unlikely.}. In fact, judging by the absence of a visible residual wobble of Eros, there may be a possibility that 
the inelastic relaxation is even faster a process than our study has shown. Some probable
physical reasons for this are discussed in (Efroimsky 2001).

\section{~~Missions ~to ~Comets~~~~~~~~~~~~~~~~~~~~~~~~~~~~~~~~~~~~~~~~~~~~~~~~~~~~~~~~~~~~~~~~~~~~~~~}

The cometary wobble is caused by tidal forces or by outgassing 
(Samarasinha \& Belton 1995). Preliminary estimates (Efroimsky 2001) show
that precession damping of a comet may be registered within a one-year time 
span or so, provided the best available spacecraft-based devices are 
used and the comet's spin state is not too close to the separatrix. Several
missions to comets, including wobbling comets, are currently being planned. 
CONTOUR spacecraft, to be launched in 2002, will fly by Encke, 
Schwassmann-Wachmann 3, and  d'Arrest, but the encounters will be brief. 
The mission will observe the spin states, but not 
their evolution. Another mission, Stardust, that is to visit comet P/Wild 2,
will also be a fly-by one. Spececraft Deep Impact will approach comet 
9P Tempel 1 on the 3 of July 2005, and will shoot a 500-kg impactor at 10 km/s
speed, to blast a crater into the nucleus, to reveal its interior. Sadly,
though, this encounter too will be short.

Rosetta orbiter is designed to approach 46/P Wirtanen in 2001 and to 
escort it for about 2.5 years (Hubert \& Schwehm 1991). Wirtanen is a wobbling 
comet (Samarasinha, Mueller \& Belton 1996), (Rickman \& Jorda 1998), and one 
of the planned experiments is observation of its rotation state, to be carried 
out by the OSIRIS camera (Thomas et al 1998). The mission will start about 1.5 
year before the perihelion, but the spin state will be observed only once,
about a 
year before the perihelion, at a heliocentric distance of 3.5 AU. Three 
months later the comet should come within 3 AU which is to be a crucial 
threshold for its jetting activity: at this distance outgassing of water will 
begin. The strongest non-gravitational torques emerge while 
the comet is within this distance from the Sun. It is predominantly during 
this period that wobble is instigated. Hence, it may be good to expand the 
schedule of the spin-state observations: along with the measurement currently 
planned for 3.5 AU before perihelion, another measurement, at a similar 
heliocentrical distance after the perihelion, would be useful. The comparison 
of these observations will provide information of the precession increase 
during the time spent by the comet within the close proximity to the Sun. 
Unfortunately, the Rosetta programme will be over soon afterwards, and a third 
observation at a larger distance will be impossible. The third observation 
performed well outside the 3 AU region might reveal the wobble-damping rate, 
and thereby provide valuable information about the composition and inner 
structure of the comet nucleus. It would be most desirable to perform such a 
three-step observation by the future escort missions to comets. What are the 
chances of success of such an experiment? On the one hand, the 
torques will not be fully eliminated after the comet leaves the 3 AU
proximity of the Sun; though the outgassing of water will cease, some 
faint sublimation of more volatile species (like CO, CO$_2$, CH$_3$OH) will 
persist for long. On the other hand,  the damping rate is high enough 
(Efroimsky 2001), and a comparison of two observations separated by 
about a year will have a very good chance of registering relaxation (provided 
the spin state is not too close to the separatrix). 

\section{~~Missions ~to ~Asteroids~~~~~~~~~~~~~~~~~~~~~~~~~~~~~~~~~~~~~~~~~~~~~~~~~~~~~~~~~~~~~~~~~~~~~}   

Asteroids wobble either after tidal interactions (if they are  
planet-crossers), or after impacts, or if they are  
fragments of disrupted progenitors (Asphaug \& Scheeres 1999). To register 
relaxation of a solid-rock monolith may take thousands of years (Efroimsky 
2001). However, if the body is loosely connected (Harris 1998, 
Asphaug et al 1998) the inelastic dissipation in it will be
several orders faster and, appropriately, its relaxation rate will be several orders higher. Relaxation of rubble-pile planet-crossers 
may be observed through the same three-step scheme: observe the spin state shortly before the tidal interaction, then shortly afterwards, and 
then after a longer interval.

In the long run, our efforts in observation of excited spin states should target the
mainbelt. As well known, it is the perturbations in the belt that drive 
some asteroids to planet-crossing orbits. Hence the necessity of knowing the 
impact frequency. To know it, one has to collect the 
statistics of precessing rotators, on the one hand, and to have a 
reliable estimate of their relaxation rate, on the other hand.

\section{~~Conclusions~~~~~~~~~~~~~~~~~~~~~~~~~~~~~~~~~~~~~~~~~~~~~~~~~~~~~~~~~~~~~~~~~~~~~~~~~~~~}

We may be very close to observation of the relaxational dynamics of wobbling 
small Solar System bodies, dynamics that may say a lot about their structure 
and composition and also about their recent histories of impacts and tidal 
interactions. Monitoring of a wobbling comet during about a year after it leaves
the 3 AU zone will, most probably, enable us to register its precession relaxation. 

\pagebreak

\end{document}